\documentclass[twocolumn,showpacs,preprintnumbers,pra]{revtex4}
\pdfoutput=1
\usepackage{amsfonts}
\usepackage{color}
\usepackage{amsmath}
\usepackage{amssymb}
\usepackage{graphicx}
\usepackage{natbib}

\begin{document}

\title{Simultaneous vibrational resonance in the amplitude and phase
quadratures of an optical field based on Kerr nonlinearity}
\author{Yinuo Wang, Shan Wu, Cuicui Li, Zhenglu Duan, Min Xie, and Bixuan Fan%
}
\email{fanbixuan@jxnu.edu.cn}
\affiliation{College of Physics and Communication Electronics, Jiangxi Normal University,
Nanchang 330022, China}

\begin{abstract}
Vibrational resonance (VR) is a nonlinear phenomenon in which the system
response to a weak signal can be resonantly enhanced by applying a
high-frequency modulation signal with an appropriate amplitude. The majority of
VR research has focused on amplifying the amplitude or intensity of the
system response to a weak signal, whereas the study of the phase
information of system responses in VR remains limited. Here, we investigate the VR phenomena
in both amplitude and phase quadratures of an optical field in a Kerr
nonlinear cavity driven by a near-resonant weak signal and a far-detuned
modulation signal. Analytical and numerical results demonstrated that the
resonant enhancement in the amplitude and phase quadratures of the system
response to a weak signal simultaneously occurs as the amplitude of the
modulation signal is varied. There is a linear relation between the
amplitude and frequency of the modulation signal for achieving an optimal VR
effect. Furthermore, we generalized our study to investigate the
quadrature at an arbitrary phase and determined that the VR enhancement
sensitively depends on the phase. Our findings not only broaden the scope
of VR research by incorporating phase information but also introduces an
 approach for amplifying an optical field by manipulating another
optical field.
\end{abstract}

\maketitle

\section{Introduction}

Resonance refers to the phenomenon by which the amplitude of a physical system
at a certain frequency overwhelms that at other frequencies, which
commonly occurs in nature, and can be observed or applied in nearly all
branches of physics, as well as in many interdisciplinary and engineering
fields.  Stochastic resonance
(SR) is a widely-studied phenomena \cite%
{benzi1981mechanism,benzi1982stochastic,benzi1983theory,RevModPhys.70.223,Thomas_Wellens_2004}%
, in which noise is utilized to amplify the response of a bistable system to
a weak input signal and the optimal amplification or the resonance occurs
when the noise-induced average transition rate matches the frequency of the
weak signal. SR was first proposed by Benzi et al. in the study
of climate change as an enhancement of the response of bistable systems to
weak deterministic signals \cite%
{benzi1981mechanism,benzi1982stochastic,benzi1983theory}. Over time, the
study of SR has extended to the various related areas, such as coherence
resonance \cite{CR0,CR1,CR2}, resonant activation \cite%
{doering1992resonant,marchi1996resonant,ghosh2006parametric}, noise-induced
stability \cite{PhysRevLett.76.563,zeng2015noise,e17052862}, noise-induced
pattern formation \cite{sanz2001turing,pattern,das2013dichotomous}, and
noise-enhanced temporal regularity \cite{Yu_18,PhysRevA.107.023708}.

A noteworthy analogy to SR is vibrational resonance (VR) \cite%
{landa2000vibrational,jeyakumari2009single,yao2011frequency}, which occurs
when a high-frequency periodic signal replaces noise in SR to amplify the
response of a nonlinear system to a weak signal. VR was first numerically
observed by Landa and McClintock \cite{landa2000vibrational}, which was then
theoretically \cite%
{gitterman2001bistable,zaikin2002vibrational,blekhman2004conjugate,ichiki2012linear}
and experimentally \cite%
{chizhevsky2003experimental,baltanas2003experimental,chizhevsky2006experimental,chizhevsky2014vibrational}
demonstrated in a variety of systems. The study of conventional VR has been extended to a
number of variations, that is, aperiodic VR \cite%
{PhysRevE.77.051126,jia2018improving}, ghost VR \cite%
{rajamani2014ghost,usama2021vibrational}, nonlinear VR \cite%
{ghosh2013nonlinear,mbong2018controllable}, entropic VR \cite%
{PhysRevE.102.012149,2021Entropic}, logical VR \cite%
{gui2020enhanced,huang2023logical}, and vibrational anti-resonance \cite%
{sarkar2019vibrational}. Owing to the deterministic and controllable nature of
VR, it has demonstrated significant potential in several research fields, including weak
fault detection \cite{XIAO2019490,xiao2021adaptive}, weak signal
amplification \cite{VR_Weak,NanoLett_VR}, investigating atmospheric
disturbance phenomena \cite{jeevarekha2016nonlinear}, and bioinformatics
\cite%
{deng2010vibrational,wang2014vibrational,ge2020vibrational,fu2022reentrance}.

Most of the earlier research regarding VR has focused on the enhancement of the system
response to a weak signal, whereas studies regarding the phase
information of the system response in the VR phenomena \cite%
{sarkar2019vibrational} remain limited. In \cite{sarkar2019vibrational}, Sarkar and Ray
theoretically studied the phase variation of the system response in the
vibrational antiresonance phenomenon and demonstrated that a large phase
shift was induced by varying the amplitude of a high-frequency field. To the best of our knowledge, the phase properties of the system response to a weak signal in
conventional VR has not been investigated thus far. Therefore, in this study,
we investigate the VR phenomena in both the amplitude and phase quadratures of
an optical signal in the context of a driven single-mode optical cavity
containing a Kerr medium. Owing to the self-Kerr interaction of the cavity
field, the amplitude and phase quadratures are nonlinearly coupled,
providing the basis for studying the VR behaviors in both quadratures. By
directly separating the fast and slow motion, we
derived the approximate analytical expressions of the response amplitudes for
both quadratures. The response amplitude is a typical quantity that can be used to
characterize the VR phenomena, which measures the amplitude of the system
response at the frequency of the weak signal to be amplified. The results demonstrated
that VR simultaneously occurred in the amplitude and phase quadratures as we
varied the amplitude of the modulation signal. We also performed numerical
simulations to verify the analytical results and demonstrated the system dynamics.
The numerical results were qualitatively consistent with the
analytical results. Notably, the amplitude of the phase quadrature was
highly sensitive to the amplitude of the modulation signal in a certain
regime, demonstrating the potential to be applied for precision measurements. In
addition, the numerical results demonstrated that there was a linear relationship between
the modulation amplitude and frequency for the optimal condition of the
signal amplification.

In the experiments, the quadrature information of an optical field can be
obtained using the standard homodyne detection technique. By adjusting the
phase of the local oscillator in the homodyne detection process, the information for the desired quadrature of the system output
signal can be acquired. For example, the information of the amplitude and phase quadratures
discussed above can be obtained by selecting a phase of zero and $\pi/2$,
respectively. To clarity the function of the phase in VR, we numerically
investigated the dependence of the response amplitude on the phase of the
local oscillator, and the results demonstrated that the response amplitude
experienced sine-like oscillations and its maximal value did not correspond
to zero or $\pi/2$. Our findings encourage the further investigation of
the interplay between VR and the optical phase information, and provide a
theoretical guidance for simultaneously amplifying the amplitude and phase
quadratures of system responses to a weak optical field by applying
another optical field. This may have potential for applications in optical signal
detection and energy transfers between optical fields with different
frequencies.

In Section \ref{sec2}, we
introduce the physical model and provide the theoretical formulation. The main results are presented in
Section \ref{sec3}, including the VR phenomena
in the amplitude and phase quadratures of the system response, and the
dependence of VR behaviors on the system parameters. Finally, Section \ref{sec4}
concludes our study.

\section{Model and theoretical analyses}

\label{sec2}
\begin{figure}[tbp]
\centering
\includegraphics[width=3.5in]{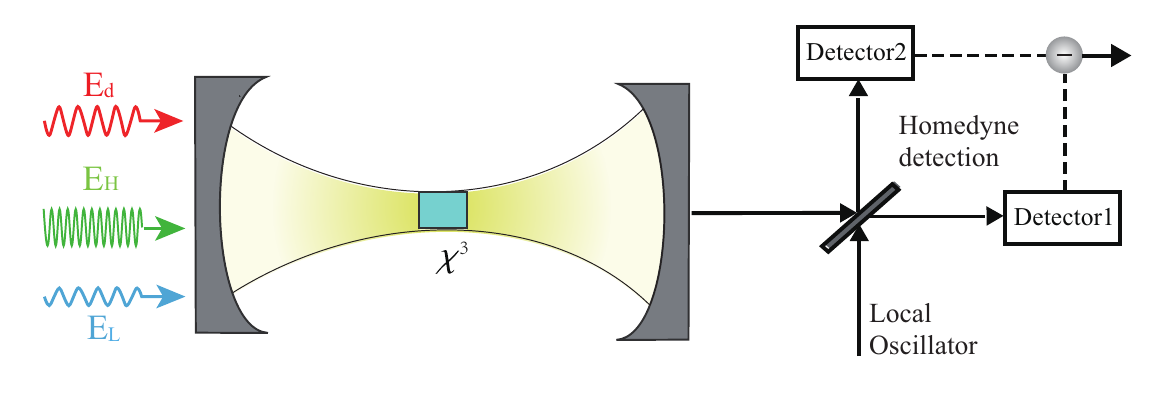}\newline
\caption{(Color online) Sketch of our model. A single-model optical cavity
containing a third-order nonlinear Kerr medium is driven by a driving field, $%
E_d$, and two signal fields, $E_H$ and $E_L$. $E_L$ is a weak signal to be
amplified and $E_H$ is a modulation signal which is far detuned from the
cavity resonance frequency. The output field of the cavity is detected
via the homodyne detection.}
\label{fig1}
\end{figure}

As shown in Fig.1, the model being considered is an anharmonic
optical oscillator with a self-Kerr interaction, that is, a single-mode optical
cavity containing a nonlinear Kerr medium. The cavity mode is driven by
three fields with different frequencies, that is, one driving field ($E_c$) and two
signal fields($E_L$ and $E_H$). The Hamiltonian describing the system is
expressed by ($\hbar =1$) as follows:
\begin{eqnarray}
\hat{H} &=&\omega _{c}\hat{a}^{\dag }\hat{a}+\chi (\hat{a}^{\dag }\hat{a}%
)^{2}+iE_{d}(e^{-i\omega _{d}t}\hat{a}^{\dagger }-e^{i\omega _{d}t}\hat{a})
\\
&+&iE_{L}(e^{-i\omega _{L}t}\hat{a}^{\dagger }-e^{i\omega _{L}t}\hat{a}%
)+iE_{H}(e^{-i\omega _{H}t}\hat{a}^{\dagger }-e^{i\omega _{H}t}\hat{a}).
\notag
\end{eqnarray}%
Here, the cavity mode with the resonance frequency $\omega _{c}$ is represented
By the annihilation operator $\hat{a}$. $E_{d}$ and $\omega _{d}$ are the
amplitude and frequency of the driving field, $E_{L}$ and $\omega _{L}$ are
the amplitude and frequency of the low-frequency weak signal, and $E_{H}$
and $\omega _{H}$ are the amplitude and frequency of the high-frequency
modulation signal. $\chi $ is the third-order nonlinear coefficient of the
Kerr medium. After rotating with respect to the driving frequency $\omega
_{d}$, the Hamiltonian [Eq.(1)] becomes the following:
\begin{eqnarray}
\hat{H} &=&\Delta \hat{a}^{\dag }\hat{a}+\chi (\hat{a}^{\dag }\hat{a}%
)^{2}+iE_{d}(\hat{a}^{\dagger }-\hat{a}) \\
&+&iE_{L}(e^{-i\Delta _{L}t}\hat{a}^{\dagger }-e^{i\Delta _{L}t}\hat{a}%
)+iE_{H}(e^{-i\Delta _{H}t}\hat{a}^{\dagger }-e^{i\Delta _{H}t}\hat{a}),
\notag
\end{eqnarray}%
where the detunings are defined as $\Delta =\omega _{c}-\omega _{d},$ $%
\Delta _{L}=\omega _{L}-\omega _{d} $, and $\Delta _{H}=\omega _{H}-\omega
_{d}.$ To satisfy the condition of VR, we assume that the two
signals have distinguishing frequencies, that is, $\Delta _{H}\gg \Delta _{L}$.
The corresponding master equation governing the dynamics of the system with
the inclusion of the dissipation induced by the interaction with the
environment is given by the following:%
\begin{equation}
\dot{\rho}=i[\rho ,\hat{H}]+\kappa \mathcal{D}[\hat{a}]\rho ,
\end{equation}%
where $\kappa $ is the decay rate of the cavity mode. By using the following relationship $%
\frac{d}{dt}Tr\left( \hat{A}\rho \right) $ $=Tr\left( \hat{A}\frac{d\rho }{dt%
}\right),$ we can derive the equation of motion for the mean amplitude of
the cavity field as follows:
\begin{eqnarray}
\frac{d\left\langle \hat{a}\right\rangle }{dt} &=&-i(\Delta +\chi
)\left\langle \hat{a}\right\rangle -\frac{\kappa }{2}\left\langle \hat{a}%
\right\rangle -2i\chi \left\langle \hat{a}^{\dag }\hat{a}^{2}\right\rangle
+E_{d}  \notag \\
&+&E_{L}e^{-i\Delta _{L}t}+E_{H}e^{-i\Delta _{H}t}.
\end{eqnarray}

Under the condition of a weak nonlinearity and strong driving field, we
approximately factorized the correlation terms, that is, $\left\langle \hat{a}%
^{\dag }\hat{a}^{2}\right\rangle \approx \left\langle \hat{a}^{\dag
}\right\rangle \left\langle \hat{a}\right\rangle^2 $, and defined a classical
variable for the mean amplitude of the cavity field $\alpha =\left\langle
\hat{a}\right\rangle$. The following was then obtained:
\begin{eqnarray}  \label{alpha}
\frac{d\alpha}{dt} &=&-\left( i\left( \Delta +\chi \right) +\kappa /2\right)
\alpha -2i\chi \left\vert \alpha \right\vert ^{2}\alpha +E_{d}  \notag \\
&+&E_{L}e^{-i\Delta _{L}t}+E_{H}e^{-i\Delta _{H}t}.
\end{eqnarray}
Note, $\alpha$ is usually a complex variable and therefore Eq.(\ref%
{alpha}) is different from the typical equation of motion in classical
nonlinear systems used for investigating the VR phenomena. To transform
Eq. (\ref{alpha}) into real variables and make the results
experimentally detectable, we defined the amplitude quadrature $x=\frac{%
\alpha +\alpha ^{\ast }}{2}$ and phase quadrature $y=\frac{\alpha
-\alpha ^{\ast }}{2i}$, thus the equations of motion of the system can be expressed as follows:%
\begin{eqnarray}  \label{x}
\frac{dx}{dt} &=&-\frac{\kappa }{2}x+\left( \Delta +\chi \right) y+2\chi
\left( x^{2}+y^{2}\right) y  \notag \\
&&+E_{d}+E_{L}\cos \Delta _{L}t+E_{H}\cos \Delta _{H}t, \\\label{y}
\frac{dy}{dt} &=&-\frac{\kappa }{2}y-(\Delta +\chi )x-2\chi \left(
x^{2}+y^{2}\right) x  \notag \\
&&-E_{L}\sin \Delta _{L}t-E_{H}\sin \Delta _{H}t.
\end{eqnarray}

Precisely solving the coupled Eqs.(\ref{x}-\ref{y}%
) is complex, as they are differential equations containing nonlinear terms and two
driving signals with distinguishing frequencies. Therefore, the
slow and fast motions were seperated \cite%
{PhysRevE.67.066119,PhysRevE.83.066205,ghosh2013nonlinear,sarkar2019vibrational}.%
Thus, the quadrature variables $x$ and $y$ were rewritten as the summation of the slow
and fast motions:%
\begin{eqnarray}  \label{x0}
x\left( t\right) &=&X\left( t\right) +\Psi _{x}\left( t,\tau =\Delta
_{H}t\right) , \\\label{y0}
y\left( t\right) &=&Y\left( t\right) +\Psi _{y}\left( t,\tau =\Delta
_{H}t\right).
\end{eqnarray}%
Here, $X\left( t\right) $ and $Y\left( t\right) $ are the variables
characterizing the slow-motion components of the system response caused by
$F_L$, whereas $\Psi _{x}\left( t,\tau \right) $ and $\Psi _{y}\left( t,\tau
\right) $ are variables characterizing the fast-motion components caused by $%
F_H$, which satisfy
\begin{eqnarray}  \label{Psi_x}
\left\langle \Psi _{x}\left( t,\tau \right) \right\rangle &=&\int_{0}^{2\pi
}\Psi _{x}\left( t,\tau \right) d\tau =0, \\\label{Psi_y}
\left\langle \Psi _{y}\left( t,\tau \right) \right\rangle &=&\int_{0}^{2\pi
}\Psi _{y}\left( t,\tau \right) d\tau =0.
\end{eqnarray}

By substituting Eqs.(\ref{x0}-\ref{y0}) into Eqs.(\ref{x}-\ref{y}) and
averaging over one period of the fast motion $\tau$, the following is obtained:
\begin{eqnarray}  \label{dX}
\frac{dX}{dt} &=&-\frac{\kappa }{2}X+\left( \Delta +\chi \right)
Y+E_{d}+E_{L}\cos \Delta _{L}t  \notag \\
&&+2\chi (X^{2}Y+2X\left\langle \Psi _{x}\Psi _{y}\right\rangle
+Y^{3}+\allowbreak Y\left\langle \Psi _{x}^{2}\right\rangle  \notag \\
&&+3Y\left\langle \Psi _{y}^{2}\right\rangle +\left\langle \Psi _{x}^{2}\Psi
_{y}\right\rangle +\left\langle \Psi _{y}^{3}\right\rangle ), \\\label{dY}
\frac{dY}{dt} &=&-\frac{\kappa }{2}Y-(\Delta +\chi )X-E_{L}\sin \Delta _{L}t
\notag \\
&&-2\chi (X^{3}+XY^{2}+3X\left\langle \Psi _{x}^{2}\right\rangle
+X\left\langle \Psi _{y}^{2}\right\rangle  \notag \\
&&+2Y\left\langle \Psi _{x}\Psi _{y}\right\rangle +\left\langle \Psi
_{x}\Psi _{y}^{2}\right\rangle +\left\langle \Psi _{x}^{3}\right\rangle ).
\end{eqnarray}

Subsequently, the equations for the fast motion can obtained by subtracting the
slow motion Eqs.(\ref{dX}-\ref{dY}) from the original
motion Eqs. (\ref{x}-\ref{y}) as follows:
\begin{eqnarray}
\frac{d\Psi _{x}}{dt} &=&-\frac{\kappa }{2}\Psi _{x}+\left( \Delta +\chi
\right) \Psi _{y}+E_{H}\cos \Delta _{H}t \\
&&+2\chi (X^{2}\Psi _{y}+2XY\Psi _{x}+3Y^{2}\Psi _{y}+  \notag \\
&&2X\left( \Psi _{x}\Psi _{y}-\left\langle \Psi _{x}\Psi _{y}\right\rangle
\right) +\allowbreak Y\left( \Psi _{x}^{2}-\left\langle \Psi
_{x}^{2}\right\rangle \right)  \notag \\
&&+3Y\left( \Psi _{y}^{2}-\left\langle \Psi _{y}^{2}\right\rangle \right)
+\Psi _{x}^{2}\Psi _{y}-\left\langle \Psi _{x}^{2}\Psi _{y}\right\rangle
+\Psi _{y}^{3}-\left\langle \Psi _{y}^{3}\right\rangle )  \notag \\
\frac{d\Psi _{y}}{dt} &=&-\frac{\kappa }{2}\Psi _{y}-(\Delta +\chi )\Psi
_{x}-E_{H}\sin \Delta _{H}t \\
&&-2\chi (\allowbreak 3X^{2}\Psi _{x}+2X\left( \Psi _{x}^{2}-\left\langle
\Psi _{x}^{2}\right\rangle \right)  \notag \\
&&+Y^{2}\Psi _{x}+2XY\Psi _{y}+X\left( \Psi _{y}^{2}-\left\langle \Psi
_{y}^{2}\right\rangle \right)  \notag \\
&&+2Y\Psi _{x}\Psi _{y}-2Y\left\langle \Psi _{x}\Psi _{y}\right\rangle +\Psi
_{x}\Psi _{y}^{2}-\left\langle \Psi _{x}\Psi _{y}^{2}\right\rangle )  \notag
\end{eqnarray}

As $\Delta _{H}$ is assumed to be large, $\dot{\Psi}_{x},\dot{\Psi}_{y}\gg
\Psi _{x},\Psi _{y}$, we can obtain the approximate
solutions for the fast-motion variables as follows:%
\begin{eqnarray}
\Psi _{x} &\approx&\frac{E_{H}}{\Delta _{H}}\sin \Delta _{H}t, \\
\Psi _{y} &\approx&\frac{E_{H}}{\Delta _{H}}\cos \Delta _{H}t.
\end{eqnarray}%
We then obtain $\left\langle \Psi _{x}^{2}\right\rangle =\left\langle \Psi
_{y}^{2}\right\rangle =\frac{E_{H}^{2}}{2\Delta _{H}^{2}}$ and $\left\langle
\Psi _{x}\Psi _{y}\right\rangle =\left\langle \Psi _{x}^{2}\Psi
_{y}\right\rangle =\left\langle \Psi _{x}\Psi _{y}^{2}\right\rangle =0.$ By
substituting these approximate solutions for the fast motion into the
equations of motion for the slow motion [Eqs. (\ref{dX}-\ref{dY})], we can
obtain the approximate equations merely for the slow motion of the system
variables, that is:
\begin{eqnarray}  \label{slowx}
\frac{dX}{dt} &=&-\frac{\kappa }{2}X+(\Delta +\chi +\frac{4\chi E_{H}^{2}}{%
\Delta _{H}^{2}})Y  \notag \\
&&+2\chi \left( X^{2}Y+Y^{3}\right) +E_{d}+E_{L}\cos \Delta _{L}t, \\\label{slowy}
\frac{dY}{dt} &=&-\frac{\kappa }{2}Y-(\Delta +\chi +\frac{4\chi E_{H}^{2}}{%
\Delta _{H}^{2}})X  \notag \\
&&-2\chi \left( \allowbreak X^{3}+XY^{2}\right) -E_{L}\sin \Delta _{L}t.
\end{eqnarray}
Eqs.(\ref{slowx}-\ref{slowy}) demonstrate that the high-frequency
signal impacts the steady-state properties of the system by
the additional detuning applied to the cavity field.

To evaluate the system response to the weak signal $E_{L}$, we
first searched for the steady-state solution for $X$ and $Y$ in the absence of
the weak signal. The steady-state solution of the
field intensity $|\alpha|_s^2$ simply satisfies the following equation:
\begin{eqnarray}
&&4\chi^2 (|\alpha|_s^2)^3+4\chi(\Delta+\chi+4\chi
E_H^2/\Delta_H^2)(|\alpha|_s^2)^2  \notag \\
&&+((\Delta+\chi)^2+\kappa^2/4)|\alpha|_s^2-E_d^2=0,
\end{eqnarray}
which is a cubic equation of $|\alpha|_s^2$ and can be solved by a standard
formula or numerically. From Eqs.(\ref{slowx}-\ref{slowy}) and the relationship of
$|\alpha|^2_s=X^2_s+Y^2_s$, we can obtain the steady-state solution of the field
Quadratures as follows:
\begin{eqnarray}
&&Y_s=-\frac{2}{\kappa}(\Delta+\chi+2\chi|\alpha|^2_s), \\
&&X_s=\frac{E_d}{\kappa/2-Y_s(\Delta+\chi+2\chi|\alpha|^2_s)}.
\end{eqnarray}

Subsequently, we studied the deviation of $X$ and $Y$ from the steady-state solution when
a weak signal was applied. Therefore, we expressed $X$ and $Y$ as the
summation of their stable solutions and small deviation parts owing to the
signal incidence as follows:
\begin{eqnarray}
X &=&X_{s}+\delta X, \\
Y &=&Y_{s}+\delta Y.
\end{eqnarray}%
Here, $\left( \delta X,\delta Y\right) $ are the deviations of the system
responses $\left( X,Y\right) $ from one set of the steady-state solution $\left(
X_{s},Y_{s}\right).$

As the $\delta X$ and $\delta Y$ deviations were assumed to be small, the nonlinear terms were ignored and the linear equations of motion were obtained
for $\left( \delta X,\delta Y\right) $:
\begin{eqnarray}
\frac{d\delta X}{dt} &=&M_{11}\delta X+M_{12}\delta Y+E_{L}\cos \Delta _{L}t,
\\
\frac{d\delta Y}{dt} &=&M_{22}\delta Y+M_{21}\delta X-E_{L}\sin \Delta _{L}t,
\end{eqnarray}%
where $M_{11}=4\chi X_{s}Y_{s}-\frac{\kappa }{2}$, $M_{12}=\Delta +\chi +%
\frac{4\chi E_{H}^{2}}{\Delta _{H}^{2}}+2\chi X_{s}^{2}+6\chi Y_{s}^{2}$, $%
M_{21}=-(\Delta +\chi +\frac{4\chi E_{H}^{2}}{\Delta _{H}^{2}}+6\chi
X_{s}^{2}+2\chi Y_{s}^{2})$, and $M_{22}=-\left( 4\chi X_{s}\allowbreak
Y_{s}+\frac{\kappa }{2}\right) $. The
solutions can be obtained by certain mathematical derivations:
\begin{eqnarray}
\delta X &=&A\cos (\Delta _{L}t)+B\sin (\Delta _{L}t), \\
\delta Y &=&C\cos (\Delta _{L}t)+D\sin (\Delta _{L}t).
\end{eqnarray}%
Here, the coefficients are defined as follows:
\begin{eqnarray}
A &=&\frac{E_{L}}{C_{3}}[\Delta _{L}C_{2}M_{22}M_{12} \\\nonumber
&+&\Delta _{L}\left( \Delta _{L}M_{12}+C_{1}\right) \left(
C_{1}M_{11}-M_{21}M_{22}M_{12}\right) ], \\
B &=&\frac{1}{C_2}[A\left( M_{11}C_{1}-M_{21}M_{22}M_{12}\right) \\\nonumber
&+&E_{L}\left( \Delta_{L}M_{12}+C_{1}\right) ], \\
C &=&-\frac{1}{C_1}[AM_{21}M_{22}+\Delta _{L}\left( BM_{21}-E_{L}\right) ], \\
D &=&\frac{AM_{21}+CM_{22}}{\Delta _{L}},
\end{eqnarray}%
with $C_{1}= \Delta _{L}^{2}+M_{22}^{2} $, $C_{2}=\Delta
_{L}\left( M_{21}M_{12}+C_{1}\right) $, and $%
C_{3}=C_{2}M_{21}M_{22}^{2}M_{12}-C_{1}C_{2}\left( \Delta
_{L}^{2}+M_{21}M_{12}\right) -\Delta _{L}\left(
M_{11}C_{1}-M_{21}M_{22}M_{12}\right) \left(
C_{1}M_{11}-M_{21}M_{22}M_{12}\right) .$

\begin{figure*}[btp]
\centering
\includegraphics[width=7in]{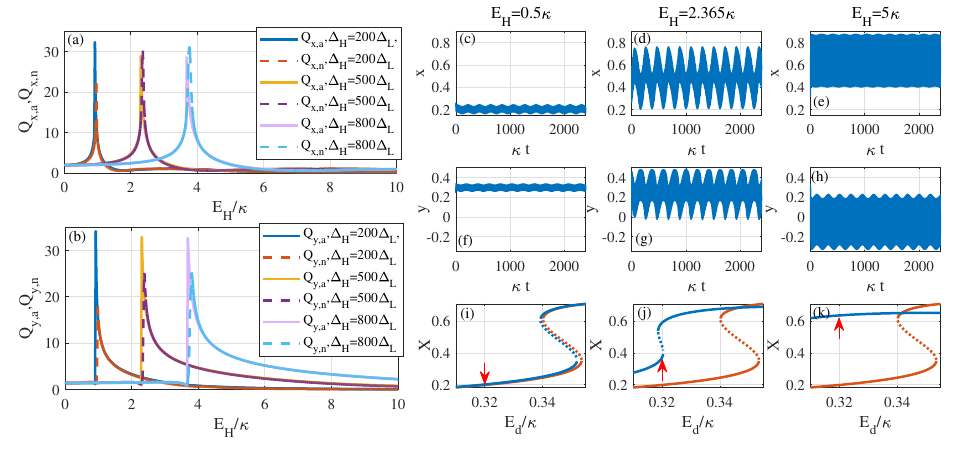}\newline
\caption{(Color online) The VR phenomena in both the amplitude and phase
quadratures of a weak optical signal by varying the amplitude of the
modulation signal $E_H$.(a-b) Comparison of the analytical and numerical
results for the response amplitudes in two quadratures as $E_H$ is varied at
three modulation frequencies ($\Delta _{H} = 200\Delta _{L}, 500\Delta _{L}$%
, and $800\Delta _{L}$). The solid curves represent the analytical results $%
Q_{x,a}$ and $Q_{y,a}$, whereas the dashed curves represent the numerical
results $Q_{x,n}$ and $Q_{y,n}$. (c-e) Time evolution of the amplitude quadrature of
the cavity field for $E_H=0.5\protect\kappa, 2.377\protect\kappa, 5\protect%
\kappa$. (f-h) Time evolution of the phase quadrature corresponding to (c-e). (i-k) Stability curves: Steady-state solution of $X$ as a function of
the driving amplitude $E_d$. As labelled by the red arrow, $%
E_d=0.32\protect\kappa$ was selected for Figs. 2(a-h). The other parameters were as
follows: $\Delta=-2\protect\kappa ,\protect\chi=\protect\kappa, E_{L}=0.004%
\protect\kappa$ and $\Delta _{L}=0.03\protect\kappa$.}
\label{fig2}
\end{figure*}

A standard measure of quantitatively characterizing the VR phenomena is the
response amplitude $Q$, which is defined as the ratio between the amplitude
of the system response at the signal frequency and amplitude of the
input signal. As both quadratures of the cavity field are involved in our
model, we evaluated the response amplitudes for the amplitude and
phase quadratures as follows:
\begin{equation}
Q_{x,a}=\frac{\sqrt{A^{2}+B^{2}}}{E_{L}},  \label{Qx_a}
\end{equation}%
\begin{equation}
Q_{y,a}=\frac{\sqrt{C^{2}+D^{2}}}{E_{L}}.  \label{Qy_a}
\end{equation}%
These approximate analytical expressions [Eqs. (\ref{Qx_a}-\ref{Qy_a})] are
the basic results obtained for analyzing the VR behavior in our system.

\section{Results and discussions}

\label{sec3}

Based on the analytical expressions [Eqs.(\ref{Qx_a})-\ref{Qy_a}], we
presented the response amplitudes $Q_{x,a}$ and $Q_{y,a}$ as a function of the
modulation amplitude $E_{H}$ for three different modulation frequencies of $%
\Delta _{H}=200\Delta _{L},500\Delta _{L},800\Delta _{L}$ [solid curves in
Figs.2(a-b)]. All the response amplitudes $Q_{x,a}$ and $%
Q_{y,a}$ apparently peak at certain values of $E_{H}$, indicating the occurrence of VR.
As the modulation frequency $\Delta _{H}$ increased, the peak
positions of $Q_{x,a}$ and $Q_{y,a}$ tended to shift to larger values of $E_{H}$%
, which is consistent with the results obtained in \cite{ghosh2013nonlinear,2021Vibrational}.
This can be explained by Eqs.(\ref{slowx}-\ref{slowy}), which indicate that the value of $%
E_{H}^{2}/\Delta _{H}^{2}$ must be maintained above a certain level to
ensure that the high frequency signal has an effective influence on the
system. In addition, the peak positions of $Q_{x,a}$ and $Q_{y,a}$ are
nearly overlapping in the axis of $E_{H}$ with the same modulation frequency $%
\Delta _{H}$, which implies that VR simultaneously occurs in two quadratures
of the system response. Note, the $Q_{y,a}$ curves
exhibit a sharp transition, which reveals the high sensitivity of $Q_{y,a}$ to
the variation of the controlling parameter $E_{H}$.

To verify the validity of the aforementioned approximated analytical results,
the equation of motion for $\alpha$ [Eq.(\ref{alpha})]
was numerically solved using the fourth-order Runge-Kutta method and the Fourier components of
the amplitude and phase quadratures at the characteristic frequency of the
weak signal ($\Delta_L$) were computed as follows:
\begin{eqnarray}
Q_{x,s} &=&\frac{2}{n\pi }\int_{0}^{nT}x(t)\sin (\Delta_L t), \\
Q_{x,c} &=&\frac{2}{n\pi }\int_{0}^{nT}x(t)\cos (\Delta_L t), \\
Q_{y,s} &=&\frac{2}{n\pi }\int_{0}^{nT}y(t)\sin (\Delta_L t) , \\
Q_{y,c} &=&\frac{2}{n\pi }\int_{0}^{nT}y(t)\cos (\Delta_L t),
\end{eqnarray}
where $T=2\pi/\Delta_L$ is the period of the weak signal $E_L$, $n$ is
the number of periods of the slow motion determined by the weak signal in the
simulation, $Q_{j,s} (j=x,y)$ and $Q_{j,c} (j=x,y)$ represent the sine and
cosine components for amplitude and phase quadratures, respectively. Subsequently,
the numerical response amplitudes for the two quadratures can be obtained as follows:
\begin{eqnarray}
Q_{x,n}&=& \frac{\sqrt{Q_{x,s}^2+Q_{x,c}^2}}{E_L} \\
Q_{y,n}&=&\frac{\sqrt{Q_{y,s}^2+Q_{y,c}^2}}{E_L}
\end{eqnarray}
The value of $Q_{j,n}$ $(j=x,y)$ is proportional to the Fourier transform
coefficient at $\omega=\Delta_L$, that is, $F_x(\omega)=\int^{+\infty}_{0}
x(t)e^{i\omega t}dt$ or $F_y(\omega)=\int^{+\infty}_{0} y(t)e^{i\omega t}dt$%
. The numerical results, $Q_{x,n}$ and $Q_{y,n}$, are indicated by the dashed
curves in Fig.2(a) By comparison, the numerical results were
apparently qualitatively consistent with the analytical results, demonstrating the
validity of the analytical calculations; however, there were certain deviations
between the two.

To understand the physical reasons behind the trends of $Q _ {x,n} $ and $Q _
{y,n}$ as $E_H$ was varied, Figs.2(c-h) present the dynamics of the
amplitude and phase quadratures at the three representative points of $E_H =
0.5\kappa, 2.365\kappa , 5\kappa$, as well as the corresponding stability curves.
For simplicity, $\Delta _ {H} = 500\Delta_L$ was considered for
example. When $E_H=0.5\kappa$, the modulation signal was weak and the system
was in the monostable state, as shown in Fig.2(i). Thus, the signals in both
quadratures oscillated around the steady-state value with a small amplitude
[Fig.2(d,g)], and the amplification of the weak signal was not significant.
When $E_H$ was increased to $2.365\kappa$, which is the optimal value for
maximizing $Q_{x,n}$ and $Q_{y,n}$ [Fig.2(a-b)], the system became bistable
[Fig.2(j)], and the system responses in both quadratures experienced
oscillations with notably larger amplitudes [Fig.2(c,f)], resulting in
significantly amplified signals at the frequency of $\Delta_L$. When the
amplitude of the high-frequency signal was further increased, that is, $E_H
=5\kappa$, the system became monostable once again [Fig.2(k)] and the
low-frequency motion was nearly buried in the strong rapid oscillations
[Fig.2(e,h)], which is consistent with the low $Q_{x,n}$ and $Q_{y,n}$ presented in Fig.2(a-b).

\begin{figure}[tbp]
\centering
\includegraphics[width=3.5in]{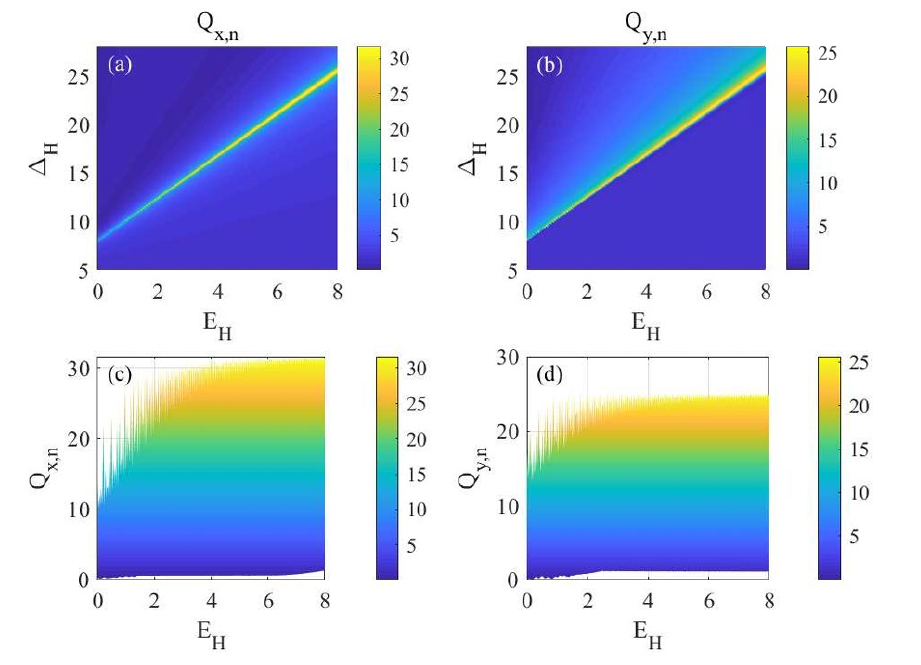}\newline
\caption{(Color online) Dependence of the response amplitudes $Q _ {x,n}
$ and $Q _ {y,n}$ on the parameters of the modulation signal ($E_H$ and $\Delta_H$). (a) $Q _ {x,n} $
in the $E_H$-$\Delta _ {H}$ plane;(b) $Q _ {y,n} $ in the $E_H$-$E_{L}$
plane. (c-d) View of the variance of $Q_{x,n}$ and $Q_{y,n}$ versus $E_H$ corresponding to (a-b). The parameters are as follows: $\Delta=-2\protect\kappa,\protect\chi=\protect%
\kappa, ,E_{d }=0.32 \protect\kappa$, $E_L=0.004\kappa$, and $\Delta _{L}=0.03\protect\kappa$. }
\label{fig3}
\end{figure}
To further investigate the detailed dependence of the system
responses on the properties of the modulation signal, the response
amplitudes of the two quadratures were plotted in the $E_H-\Delta_H$ plane, as shown in
Figs.3(a-b), which demonstrated an apparent linear relationship between $E_H$ and
$\Delta _ {H}$ for achieving resonance. These results are consistent with previous
analytical analyses, which demonstrated that the modulation signal  modifies the properties of the system
stability by the factor of $E_H^2/\Delta_H^2$. In addition, as $%
E_H$ was increased from 0 to $4\kappa$, the system responses in the two
quadratures gradually increased. However, when $E_H>4\kappa$, the system
response saturated, indicating that an extremely large $E_H$ cannot induce
a significant enhancement of the system responses and a moderate modulation
signal is adequate for achieving a good VR enhancement.

\begin{figure}[tbp]
\centering
\includegraphics[width=3.5in]{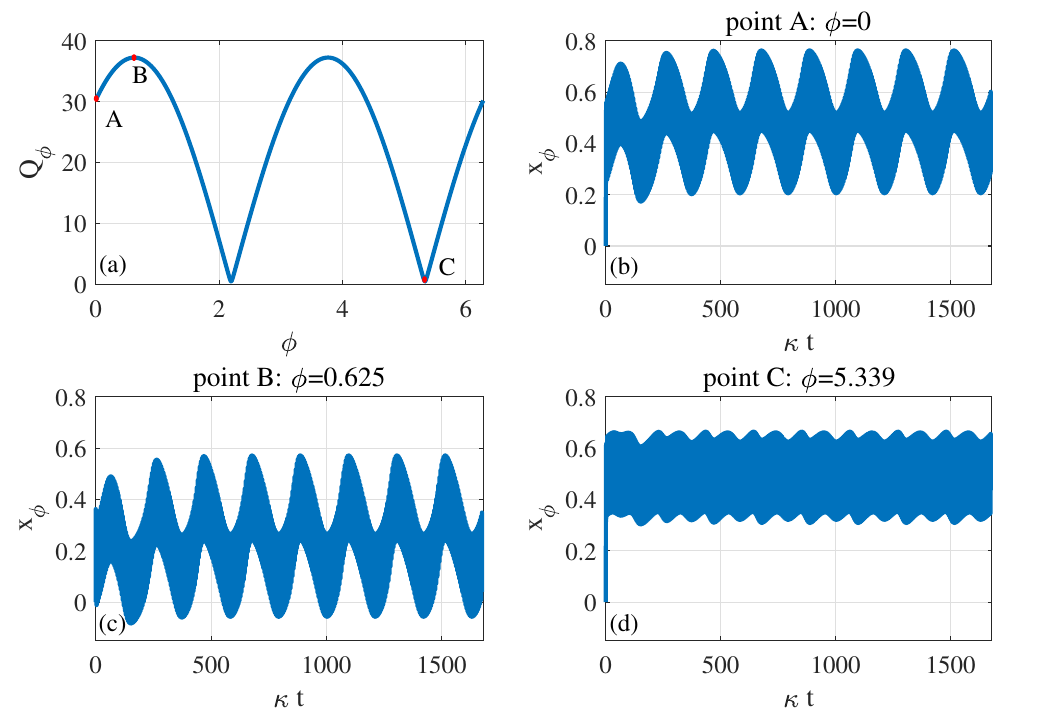}\newline
\caption{(Color online) (a) Response amplitude of an arbitrary
quadrature $x_\protect\phi$ versus the phase $\protect\phi$. (b-d) Time
evolution of $x_\protect\phi$ at three representative points including A,B, and C in
(a). Here, $E_H=2.365\protect\kappa$, $\Delta_H=500\Delta_L$, and the other
parameters are the same as those shown in Fig.3.}
\label{fig4}
\end{figure}

The output quadrature from homodyne detection apparently relies on the
phase of the local oscillator (labelled as $\phi$), that is, $x$ and $y$
presented in the preceding section corresponds to $\phi=0$ and $\phi=\pi/2$,
respectively. To generalize our results, the enhancement of
the system response at the signal frequency of $\Delta_L$ for an arbitrary
phase in the homodyne detection was studied. In this case, the homodyne signal is expressed as follows:
\begin{eqnarray}
x_{\phi,\mathrm{out}}=\sqrt{\kappa_0}x_\phi=\frac{1}{2}(\alpha_\mathrm{out}
e^{i\phi}+\alpha^*_\mathrm{out} e^{-i\phi}).
\end{eqnarray}
Here, $\alpha_\mathrm{out}$ is the amplitude of the output cavity field,
which is proportional to the intracavity amplitude $\alpha$, that is, $\alpha_%
\mathrm{out}=\sqrt{\kappa_0}\alpha$ ($\sqrt{\kappa_0}$ is the coupling
coefficient between the cavity and the homodyne detection device). Thus, we
defined the response amplitudes for the quadrature of the phase $\phi$ as follows:
\begin{eqnarray}
Q_{\phi}&=& \frac{\sqrt{Q_{\phi,s}^2+Q_{\phi,c}^2}}{E_L},
\end{eqnarray}
where the sine and cosine Fourier components at the phase $\phi$ are as follows:
\begin{eqnarray}
Q_{\phi,s} &=&\frac{2}{n\pi }\int_{0}^{nT}\frac{1}{2}(\alpha
e^{i\phi}+\alpha^* e^{-i\phi})\sin (\Delta_L t), \\
Q_{\phi,c} &=&\frac{2}{n\pi }\int_{0}^{nT}\frac{1}{2}(\alpha
e^{i\phi}+\alpha^* e^{-i\phi})\cos (\Delta_L t).
\end{eqnarray}
As demonstrated in Fig.4(a), the response amplitude $Q_\phi$ oscillated in a
sine-like form as the phase varied from $0$ to $2\pi$, indicating that the
system response was phase-sensitive. Notably, $Q_\phi$ did not peak at $%
\phi=0$ or $\phi=\pi/2$. Namely, selecting a proper phase of the
local oscillator facilitates the signal enhancement effect in VR. To
clarify the cause of variation of $Q_\phi$, the time
evolution $x_\phi$ of several representative points, including [A: $\phi=0$
(x quadrature), B: $\phi=0.625$ (one peak of the $Q_\phi$ curve), and C:$%
\phi=5.339$ (one dip of $Q_\phi$ curve)], are presented in Figs.4(b-d). At the peak, the oscillation amplitude of $x_\phi$ is apparently larger
than that of $\phi=0$, whereas at the dip, $x_\phi$ oscillates at significantly high
frequencies associated with a small amplitude of the slow variations, because
the varying phase $\phi$ corresponds to the superposition of the real and
imaginary parts of $\alpha$ with different weights. These dynamical
behaviors are consistent with the values of $Q_\phi$.

\section{Conclusion}

\label{sec4} In this study, the VR phenomenon in a Kerr nonlinear optical
cavity with multiple signals was thoroughly analyzed. Contrary to the
majority of prior research regarding VR, we incorporated the phase in the
investigation of VR. More specifically, we analytically and numerically
studied the enhancement in the amplitude and phase quadratures of the
system response to a weak low-frequency optical signal by manipulating a
high-frequency optical signal. We clarified the optimal parameter regimes
required to achieve an effective VR effect. In addition, we generalized
our study to an arbitrary quadrature of the system response and found that
the system response sensitively relies on the phase of the local oscillator in
the homodyne detection.  Our study provides a better understanding of the VR
mechanism as well as a theoretical guidance for amplifying a weak optical
signal by controlling another optical field based on the Kerr
nonlinearity.


\end{document}